\documentclass[10pt,conference]{IEEEtran}
\IEEEoverridecommandlockouts

\usepackage{graphicx}
\usepackage[hyphens]{url}
\usepackage{amsfonts}
\usepackage{booktabs}
\usepackage{romannum}
\usepackage{amsmath}
\usepackage{bm}
\usepackage{color}

\ifCLASSOPTIONcompsoc
    \usepackage[nocompress]{cite}
\else
    \usepackage{cite}
\fi
\def\BibTeX{{\rm B\kern-.05em{\sc i\kern-.025em b}\kern-.08em
    T\kern-.1667em\lower.7ex\hbox{E}\kern-.125emX}}

\begin{document}

\title{
\textsc{SmartIntentNN}:\\Towards Smart Contract Intent Detection}

\author{
    Youwei Huang$^{1}$,  Sen Fang$^{2}$, Jianwen Li$^{1, 3}$, Bin Hu$^{4}$, and Tao Zhang$^{5\ast}$
    \\
    \normalsize $^{1}$ Institute of Intelligent Computing Technology, Suzhou, CAS, China
    \\
    \normalsize $^{2}$ North Carolina State University, USA
    \\
    \normalsize $^{3}$ Beijing Normal University - Hong Kong Baptist University United International College, China
    \\
    \normalsize $^{4}$ Institute of Computing Technology, Chinese Academy of Sciences, China
    \\
    \normalsize $^{5}$ Macau University of Science and Technology, Macao SAR
    \\
    \normalsize huangyw@iict.ac.cn, tazhang@must.edu.mo
    \\
    \normalsize $^{\ast}$Corresponding author
}

\maketitle

\begin{abstract}
Smart contracts on the blockchain offer decentralized financial services but often lack robust security measures, leading to significant economic losses. While substantial research has focused on identifying vulnerabilities in smart contracts, a notable gap remains in evaluating the malicious intent behind their development. To address this, we introduce \textsc{SmartIntentNN} (Smart Contract Intent Neural Network), a deep learning-based tool designed to automate the detection of developers' intent in smart contracts. Our approach integrates a Universal Sentence Encoder for contextual representation of smart contract code, employs a K-means clustering algorithm to highlight intent-related code features, and utilizes a bidirectional LSTM-based multi-label classification network to predict ten distinct categories of unsafe intent. Evaluations on 10,000 real-world smart contracts demonstrate that \textsc{SmartIntentNN} surpasses all baselines, achieving an F1-score of 0.8633.

A demo video is available at \url{https://youtu.be/otT0fDYjwK8}.
\end{abstract}


\section{Introduction}
A smart contract is a type of computer program and transaction protocol, engineered to execute, control, or document legally binding events and actions automatically according to the stipulations of a contract or agreement~\cite{buterin2014next}. Users generally interact with smart contracts by initiating transactions to invoke various functions. From a programming standpoint, current research on smart contract security predominantly focuses on identifying vulnerabilities and bugs. However, these contracts, while serving as transaction protocols, can be compromised by developers with malicious intent, leading to substantial financial losses.

\begin{figure}[ht]
    \centering\includegraphics[width=1\linewidth]{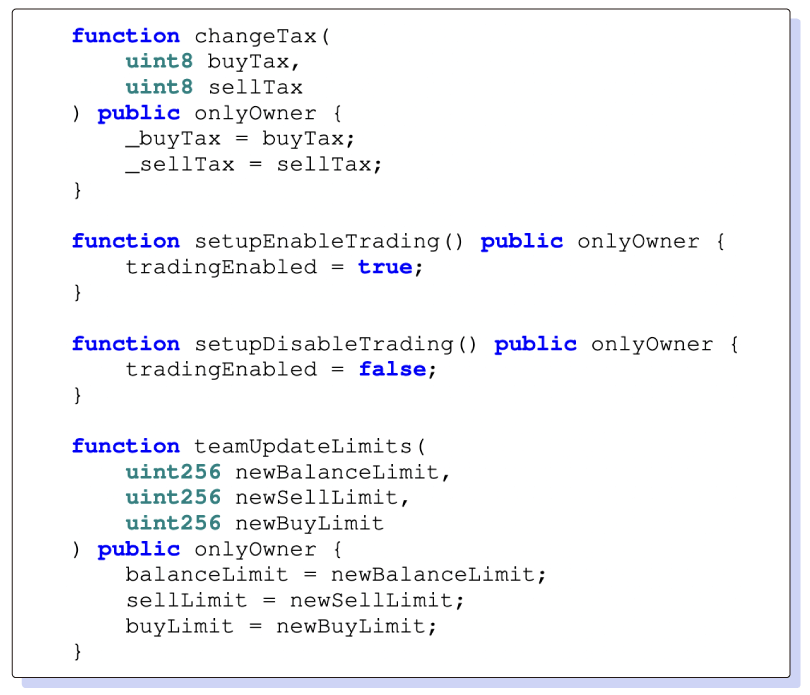}
    \caption{
        Examples of a smart contract with negative intents.
        BSC address: 0xDDa7f9273a092655a1cF077FF0155d64000ccE2A.
    }
    \label{intentalriskem}
\end{figure}

Figure \ref{intentalriskem} illustrates several samples of suspicious intent in a real smart contract. All functions share a modifier \textbf{\textit{onlyOwner}}, indicating control by a specific account. For instance, the \textbf{\textit{onlyOwner}} modifier in the \textit{changeTax} function restricts tax fee changes to the development team, while \textit{teamUpdateLimits} allows modifications to transaction limits. 
Other functions exhibit even more detrimental development intent, permitting the owner to enable or disable the trading function within the smart contract. 
Unfortunately, current research lacks effective methods for detecting developers' intent in smart contracts, and manual detection is both time-consuming and costly.

To address the gap in detecting intent in smart contracts, we propose \textsc{SmartIntentNN}, an automated deep learning-based tool designed for smart contract intent detection. It integrates a Universal Sentence Encoder (USE)~\cite{cer2018universal} to generate contextual embeddings~\cite{mikolov2013distributed} of smart contracts, a K-means clustering model~\cite{krishna1999genetic} to identify and highlight intent-related features, and a bidirectional long short-term memory (BiLSTM)~\cite{hochreiter1997long, sutskever2014sequence} multi-label classification network to predict intents in smart contracts. Trained on 10,000 smart contracts and evaluated on another 10,000 distinct contracts, this tool surpasses all baselines, achieving an \textit{F1-score} of 0.8633.

\textbf{Our contributions are as follows:}
\begin{itemize}
\item We present the first work on detecting smart contract development intent using deep learning techniques.
\item We compile an extensive dataset of over 40,000 smart contracts, labeled with ten categories of intent.
\item We open-source the code, dataset, documentation, and models at \textcolor{blue}{\url{https://github.com/web3se-lab/web3-sekit}}.
\end{itemize}

\section{Dataset}
Since \textsc{SmartIntentNN} is implemented with a deep neural network (DNN), we have amassed a dataset of over $40,000$ smart contracts sourced from the Binance Smart Chain (BSC) explorer\footnote{\url{https://bscscan.com}}. These contracts have been labeled with ten types of intent at the \textit{function} code level. The process involved downloading open-source smart contracts, merging those spanning multiple files, and removing redundant and extraneous code fragments. Finally, we extracted the \textit{function} level code snippets from these contracts.

\subsection{Intent Labels}
We categorized the smart contracts in our dataset into ten common intent categories:

\begin{itemize}
    \item[1]{
        \textbf{Fee}: Arbitrarily changes transaction fees, transferring them to specified wallet addresses.
    }
    \item[2]{
        \textbf{DisableTrading}: Enables or disables trading actions on a smart contract.
    }
    \item[3]{
        \textbf{Blacklist}: Restricts designated users' activities, potentially infringing on fair trade rights.
    }
    \item[4]{
        \textbf{Reflection}: Redistributes taxes from transactions to holders based on their holdings, attracting users to buy native tokens.
    }
    \item[5]{
        \textbf{MaxTX}: Limits the maximum number or volume of transactions.
    }
    \item[6]{
        \textbf{Mint}: Issues new tokens, either unlimited or controlled.
    }
    \item[7]{
        \textbf{Honeypot}: Traps user-provided funds under the guise of leaking funds.
    }
    \item[8]{
        \textbf{Reward}: Rewards users with crypto assets to encourage token use, despite possible lack of value.
    }
    \item[9]{
        \textbf{Rebase}: Adjusts token supply algorithmically to control price.
    }
    \item[10]{
        \textbf{MaxSell}: Limits specified users' selling times or amounts to lock liquidity.
    }
\end{itemize}

The sources of these labels include contributions from StaySafu\footnote{https://www.staysafu.org} as well as insights from decentralized application developers and auditors.

\subsection{Input Extraction}
Smart contract source code on BSC can be published either as single-file contracts with merged \textit{imports} or as multiple-file contracts. We consolidate multiple files into a single document.

We remove \textit{pragma} (Solidity compiler version specification), \textit{import} statements, and \textit{comments} as they do not affect intent expression. For multi-file contracts, \textit{import} statements become redundant after merging.

Due to the nature of smart contracts as computer code, direct input into a neural network is impractical. Instead, we use regular expressions to extract \textit{contract}-level and \textit{function}-level code. The \textit{function} code, denoted as $\mathcal{F}$, is used for model training and evaluation.

\section{\textsc{Implementation}}
The implementation of \textsc{SmartIntentNN} encompasses three primary stages: smart contract embedding, intent highlighting, and multi-label classification learning.

\subsection{Smart Contract Embedding}
To embed the context of \textit{functions}, we employ the Universal Sentence Encoder. This embedding process is denoted as $\Phi\left(\mathcal{F}\right):\mathcal{F}\rightarrow\bm{f}$, where $\Phi$ represents the contextual encoder, and $\mathcal{F}$ denotes the \textit{function} context. The output is a vector $\bm{f}$, which serves as the embedding of the \textit{function} $\mathcal{F}$.

This embedding process is applied to each \textit{function} within a smart contract. The resultant embeddings, denoted as $\bm{f}$, are aggregated into a matrix $\bm{X}$, which represents the entire smart contract. Specifically, $\bm{X} \in \mathbb{R}^{n \times m}$, where $n$ corresponds to the number of \textit{functions} in the smart contract, and $m$ represents the embedding dimension.

\subsection{Intent Highlight}
Although it is feasible to directly input $\bm{X}$ into a DNN, not all \textit{functions} are relevant to the developer’s intent. Therefore, we implement an intent highlight model to extract intent-related \textit{functions} in a smart contract. The highlighting process, denoted as $\mathrm{H}\left(\bm{X}\right):\bm{X}\rightarrow \bm{X'}$, utilizes an unsupervised model $\mathrm{H}$ to produce intent-highlighted data $\bm{X'}$.

We commence the process by training a K-means clustering model to evaluate the intent strength of each \textit{function} in randomly selecting $1,500$ smart contracts. Our experiments reveal that $19$ \textit{functions} exhibit frequencies greater than $0.75$, indicating common usage among developers. Detailed analysis suggests that these code snippets often originate from public libraries or are sections with high reuse frequency, potentially indicating a weaker developer intent. Conversely, less frequent \textit{functions} tend to express specific and strong developer intent.

To identify \textit{functions} that are significantly distant in spatial distribution from these 19 frequently occurring \textit{functions}, we initially set the number of clusters $k$ to 19 and then conducted a maximum of 80 iterations of K-means clustering training.
To compare document similarities, we compute the \textit{cosine distance} between their embedding vectors~\cite{rahutomo2012semantic, gu2018deep}. Formula~\ref{formula_similarity} defines the cosine similarity between two \textit{functions} (A and B), derived from the cosine of $\bm{f^A}$ and $\bm{f^B}$. We then transform the cosine similarity into cosine distance as defined by Formula~\ref{formula_distance}.

\begin{equation}
    \cos\left\langle\bm{f^A},\bm{f^B}\right\rangle=\frac{\bm{f^A}\cdot \bm{f^B}}{\left\|\bm{f^A}\right\|\left\|\bm{f^B}\right\|}
    \label{formula_similarity}
\end{equation}
\begin{equation}
    \mathrm{D}\left(\bm{f^A},\bm{f^B}\right)=1-\cos\left\langle\bm{f^A},\bm{f^B}\right\rangle
    \label{formula_distance}
\end{equation}

During training, the K-means model iteratively calculates the cosine distance between centroids and their within-cluster \textit{function} vectors, updating centroids to minimize the total within-cluster variation (TWCV).
This iterative process continues until no further significant reduction in TWCV occurs or the maximum iterations are reached.
During the training process of K-means clustering, some empty clusters or identical cluster centroids emerged, which were addressed by deleting or merging them, refining the number of clusters $k$ from $19$ to $16$.
Employing the trained K-means model, the within-cluster distance for each vector $\bm{f_i}$ can be predicted, which indicates the intent strength—\textbf{the greater the distance, the stronger the intent}.

\begin{equation}
    \bm{X'}=\mathrm{H_{\mu}}\left(\bm{X}\right)\;\mathsf{by}\;\mu\bm{f_i}\;\mathsf{if}\;\mathrm{D}\left(\bm{f_i},\bm{c_j}\right)\geq\lambda
    \label{formula_scale}
\end{equation}

In Formula~\ref{formula_scale}, the feature $\bm{f_i}$ in matrix $\bm{X}$ is scaled by the predicted within-cluster distance to generate a new matrix $\bm{X'} \in \mathbb{R}^{n \times m}$, where $i \in \{1, 2, \dots, n\}$ and $\bm{c_j}$ represents the cluster centroid, $j \in \{1, 2, \dots, 16\}$. Here, $\lambda = 0.21$ is the threshold; beyond it, $\bm{f_i}$ is scaled by a factor of $\mu = 16$, referred to as $\mathrm{H_{16}}$ in Section~\rm{V}. This process amplifies rare \textit{functions}, highlighting their significant intent contribution.

\subsection{Multi-label Classification}
In this section, we utilize a DNN model for multi-label binary classification. This model comprises three layers: an input layer, a BiLSTM layer, and a multi-label classification output layer. The matrix $\bm{X'}$ is fed into the model, which is trained by minimizing 10 combined binary cross-entropy losses corresponding to the 10 intent labels described in Section~\rm{II}.A.

The input layer processes sequences of dimensions $\mathbb{R}^{p \times m}$, where $p$ represents the number of \textit{functions} per time step, and $m$ represents the number of dimensions per \textit{function} embedding. Since the feature dimension is fixed across all embeddings, no modification to the columns of $\bm{X'}$ is necessary. It is essential to ensure that $m$ matches the features in $\bm{f_i}$. The row count of $\bm{X'}$ varies with the number of \textit{functions} in each smart contract. When $\bm{X'}$ has fewer rows than $p$, meaning $n < p$, the input layer, which also functions as a masking layer with a masking value of zero, pads the missing rows with zero vectors $\bm{0}$.

The subsequent layer is a BiLSTM that receives a matrix $\bm{X''} \in \mathbb{R}^{p \times m}$ from the input layer. Each LSTM layer comprises $p$ memory cells, totaling $2p$ cells due to the bidirectional configuration. Data is processed through the LSTM's input, forget, and output gates, capturing the semantic context of the smart contract. Let $h$ denote the number of hidden units, and use the vector $\bm{h}$ to represent the output of a cell. The forward pass generates $\bm{h^f}$, and the backward pass yields $\bm{h^b}$. The final output of the BiLSTM layer is the concatenation of these vectors, denoted as $\bm{h} = \bm{h^f} \oplus \bm{h^b}$~\cite{faith1967direct}.

\begin{equation}
    \bm{y}=\mathrm{sigmoid}\left(\bm{W}\bm{h}+\bm{b}\right)
    \label{formula_class}
\end{equation}

The output of the BiLSTM layer is ultimately fed into a multi-label classification dense layer. Formula~\ref{formula_class} performs binary classification for each intent label using the $\mathrm{sigmoid}$ function. The weight matrix $\bm{W}$ is defined as $\bm{W} \in \mathbb{R}^{l \times 2h}$, where $2h$ is the size of the input vector $\bm{h}$ and $l$ is the number of target labels. Consequently, the final output is a vector $\bm{y} = [y_1, y_2, \cdots, y_l]$, where each element represents the probability.
The intent detection for the smart contract is now complete.

\section{Application}
We developed \textsc{SmartIntentNN} using \textit{Tensorflow.js}~\cite{smilkov2019tensorflow}, creating a web-based tool accessible through any browser. Specifically, \textsc{SmartIntentNN} offers two primary modules: intent highlight and intent detection.

\subsection{Intent Highlight}

\begin{figure}[ht]
    \centering\includegraphics[width=\linewidth]{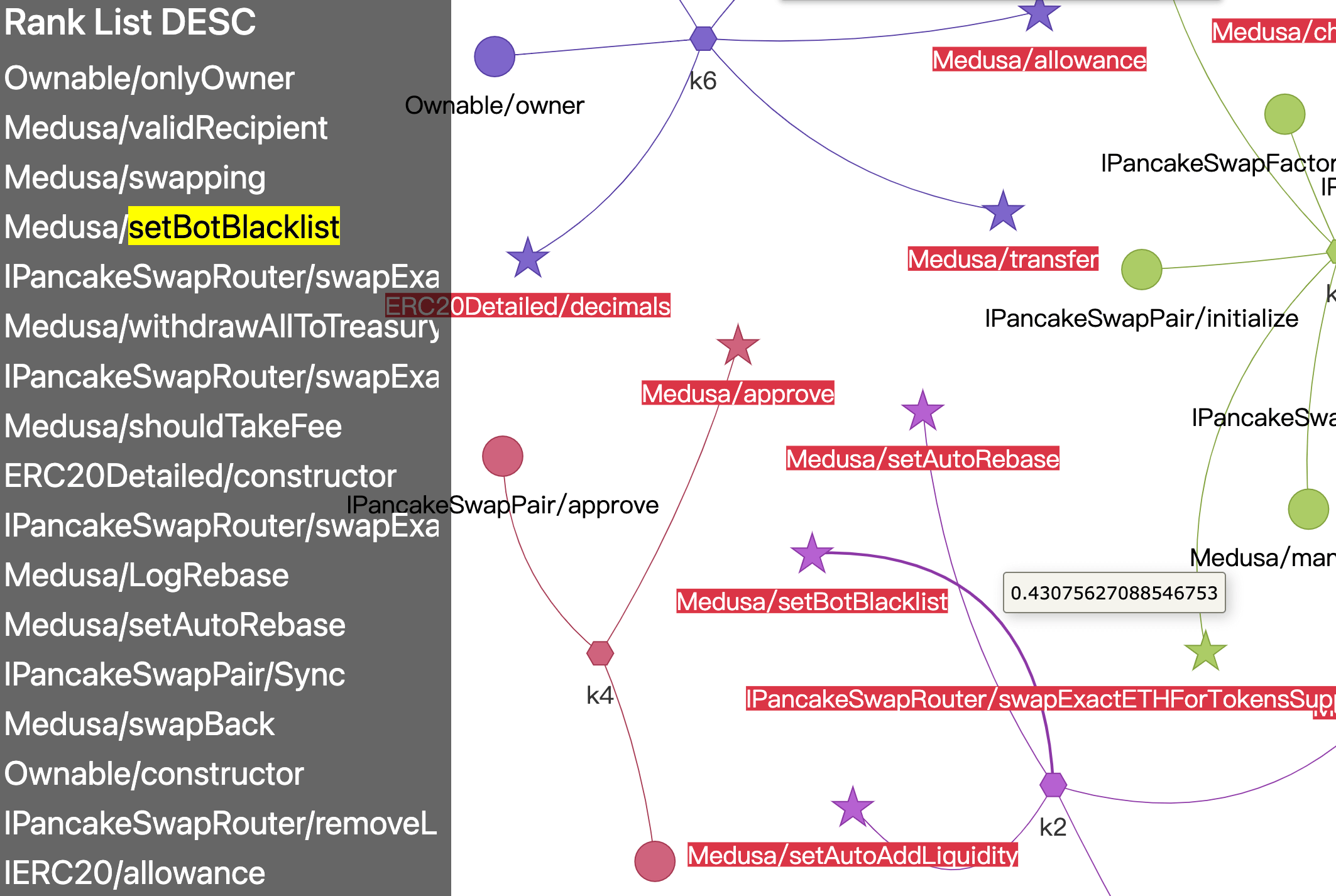}
    \caption{
        Example of intent highlighting applied to a smart contract.
        BSC address: 0xE97CBB39487a4B06D9D1dd7F17f7fBBda4c2b9c4.
    }
    \label{intent}
\end{figure}

The intent highlight feature enables users to swiftly locate \textit{functions} within smart contracts that exhibit specific, strong development intent.
In Fig.~\ref{intent}, \textit{functions} exhibiting strong intent are highlighted with a red background. 
Specifically, a hexagonal node represents the centroid of its corresponding cluster, while a circular node represents a \textit{function} with weak intent and a star represents one with strong intent.  
When an edge is focused, the distance from the centroid to the \textit{function} is displayed, indicating the strength of the intent. The user interface displays a list of \textit{functions} from a smart contract, ranked by descending intent strength on the left side.

In Fig.~\ref{intent}, several \textit{functions} are highlighted with a red background, such as \textit{setBotBlacklist} and \textit{setAutoRebase}, which indeed exhibit suspicious intent. These \textit{functions} may correspond to the intent categories of \textbf{blacklist} and \textbf{rebase} described in Section~\rm{II}.A. Non-highlighted \textit{functions} mainly include interfaces or libraries, such as those in \textit{IPancakeSwapPair}.

\subsection{Intent Detection}

\begin{figure}[ht]
    \centering
    \includegraphics[width=\linewidth]{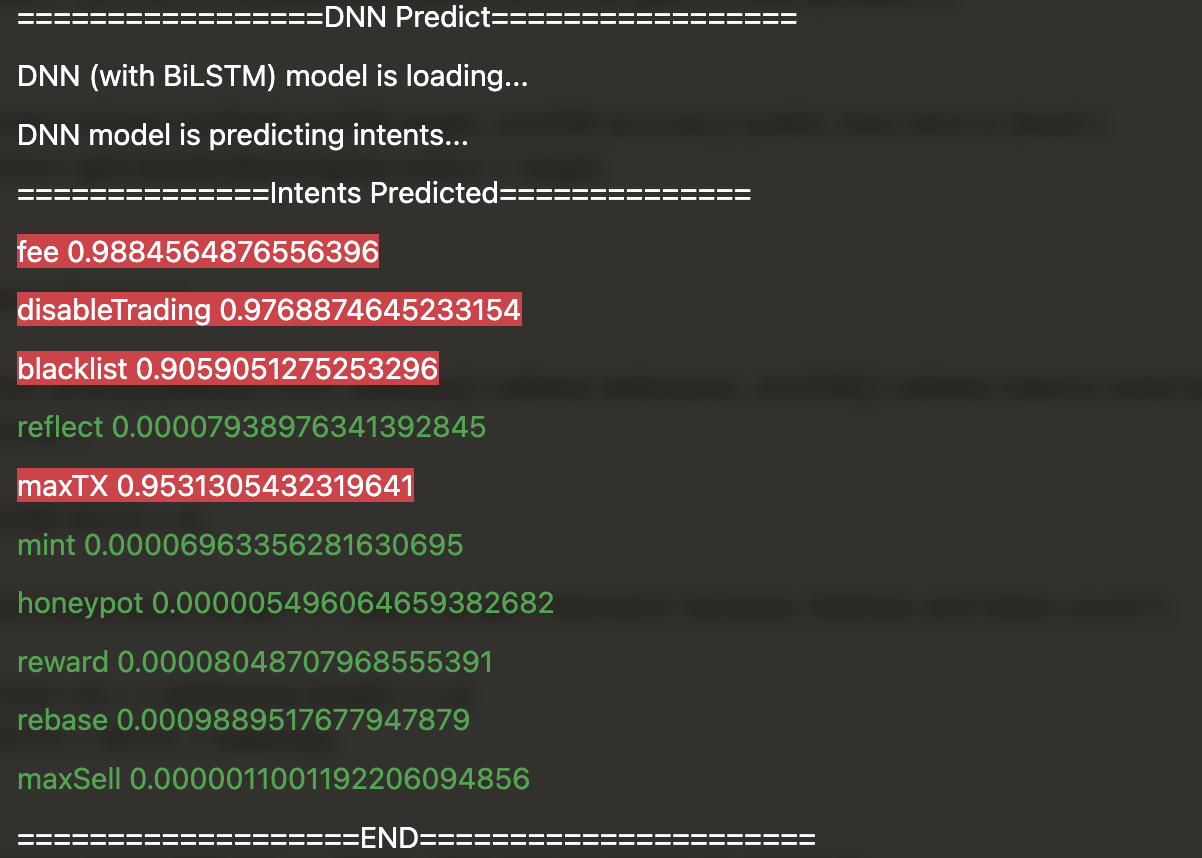}
    \caption{
        Illustration of intent detection within a smart contract.
        BSC address: 0xc4F082963E78deAaC10853a220508135505999E6.
    }
    \label{detect}
\end{figure}

Our intent detection tool features a text input area that allows users to enter or paste the source code of a smart contract. The tool employs \textsc{SmartIntentNN} to predict the intent behind various \textit{functions} in the contract. High-probability intent labels are highlighted in red, distinguishing them from low-probability labels, which are shown in green.

Figure \ref{detect} demonstrates that \textsc{SmartIntentNN} accurately identified four distinct intents within the analyzed smart contract: \textbf{fee}, \textbf{disableTrading}, \textbf{blacklist}, and \textbf{maxTX}. To validate these predictions, we performed an exhaustive manual review of the contract, confirming the existence of the aforementioned intents. Specifically, the \textbf{disableTrading} intent is controlled by the \textit{tradingOpen} variable in line $403$ and the \textit{tradingStatus} function in line $574$, while the \textbf{fee}, \textbf{maxTX}, and \textbf{blacklist} intents are encoded in the code at lines $548$ and $552$, $544$ and $630$, and $385$ and $681$, respectively.

\section{Evaluation}

To evaluate \textsc{SmartIntentNN}, we employed a confusion matrix to measure key performance metrics, including \textit{accuracy}, \textit{precision}, \textit{recall}, and \textit{F1-score}~\cite{chen2024improving}. 
%In our intent detection, successfully identifying existed intent is marked as \textit{True Positive}, no intent as \textit{True Negative}, and opposite outcomes as \textit{False Positive} and \textit{False Negative}, based on which we calculated the aforementioned four key metrics.
In our smart contract intent detection, identifying intent correctly is considered a \textit{True Positive (TP)}, correctly recognizing non-intent scenarios as \textit{True Negative (TN)}, false identifications of intent as \textit{False Positive (FP)}, and missed detections of intent as \textit{False Negative (FN)}. Based on these classifications, we further calculated \textit{accuracy}, \textit{precision}, \textit{recall}, and \textit{F1-score}.
The evaluation was conducted on a separate dataset of $10,000$ real smart contracts, which was distinct from our training dataset.

% 在我们的意图检测当中，成功检测到意图就是TP，检测到没有意图就是TN，如果与前两者相反，就是FP和FN，以此类推来计算以上的4种metrics

This research is pioneering in the field of intent detection in smart contracts and, therefore, has no prior studies for direct comparison. Consequently, we conducted a self-comparison against several established baselines, including models such as LSTM, BiLSTM, and CNN~\cite{lecun1995convolutional}. Furthermore, we benchmarked our model against popular generative large language models (LLMs) for a more comprehensive evaluation.

\begin{table}[ht]
    \centering
    \caption{Baselines Comparison}
    \begin{tabular}{@{}lcccc@{}}
        \toprule
        \textbf{Model}  & \textbf{Accuracy} & \textbf{Precision} & \textbf{Recall} & \textbf{F1-score} \\
        \midrule
        \multicolumn{5}{c}{\textbf{\textsc{SmartIntentNN} (Ablation Test)}}                            \\
        \midrule
        
        \textbf{USE-$\bm{\mathrm{H_{16}}}$-BiLSTM}  & \small $\bm{0.9647}$  & \small $\bm{0.8873}$  & \small $\bm{0.8406}$  & \small $\bm{0.8633}$     \\
        USE-$\mathrm{H_{2}}$-BiLSTM         & $0.9581$          & $0.8438$           & $0.8386$        & $0.8412$     \\    
        USE-$\mathrm{H_{16}}$-LSTM          & $0.9581$          & $0.8731$           & $0.7999$        & $0.8349$      \\
        USE-BiLSTM                          & $0.9524$          & $0.8337$           & $0.8003$        & $0.8167$      \\
        USE-LSTM                            & $0.9478$          & $0.8319$           & $0.7587$        & $0.7936$      \\
        
        \midrule
        \multicolumn{5}{c}{\textbf{Baseline Models}}                                                      \\
        \midrule
        LSTM            & $0.9172$          & $0.7725$           & $0.5973$        & $0.6737$      \\
        BiLSTM          & $0.9320$          & $0.7871$           & $0.7200$        & $0.7521$       \\
        CNN             & $0.9093$          & $0.6922$           & $0.6596$        & $0.6755$        \\
        GPT-3.5-turbo    & $0.8375$          & $0.4135$           & $0.5447$        & $0.4701$         \\
        GPT-4o-mini    & $0.7821$          & $0.3703$           & $0.9240$        & $0.5288$         \\
        \bottomrule
    \end{tabular}
    \label{tablebaseline}
\end{table}

The evaluation results presented in Table~\ref{tablebaseline} demonstrate that \textsc{SmartIntentNN} with $\mathrm{H_{16}}$ outperforms all the baselines and ablation tests, achieving an \textit{F1-score} of $0.8633$, an \textit{accuracy} of $0.9647$, a \textit{precision} of $0.8873$, and a \textit{recall} of $0.8406$. This approach markedly surpasses the baselines, with an \textit{F1-score} improvement of $28.14\%$ over LSTM, $14.79\%$ over BiLSTM, $27.80\%$ over CNN, $83.64\%$ over GPT-3.5-turbo, and $63.26\%$ over GPT-4o-mini.
We also examined two variants of the intent highlight model: $\mathrm{H_2}$ and the non-highlighted version.  
The $\mathrm{H_{2}}$ variant outperformed the non-highlighted version, with this effect being especially evident in the $\mathrm{H_{16}}$ model, which underscores the effectiveness of intent highlighting.

% 更好的原因：
% 1. 数据好and多
% 2. USE将文本转为表征，h强化意图，bilstm双向学习智能合约的上下文：最终识别出意图

% 漏洞检测、代码检测、代码生成

% GPT是一个泛化模型，没有针对我们的任务进行垂直训练

%The superior performance of \textsc{SmartIntentNN} can be attributed to the encoding of code into representations via the Universal Sentence Encoder, improved intent recognition through highlighting, and bidirectional learning of smart contract context by BiLSTM. Each of these components in the model enhances its intent detection performance.

%这种放大在h16的模型上显得尤为明显

\section{Conclusion}
In this research, we introduced \textsc{SmartIntentNN}, a novel automated tool utilizing deep learning models to detect developers' intent in smart contracts. \textsc{SmartIntentNN} incorporates a pre-trained USE model, an intent highlight model based on K-means clustering, and a DNN integrated with a BiLSTM layer. Trained on $10,000$ and evaluated on another $10,000$ distinct smart contracts, \textsc{SmartIntentNN} achieved an \textit{F1-score} of $0.8633$. The model surpasses all baselines, including the latest generative LLMs.

\bibliographystyle{IEEEtran}
\bibliography{IEEEabrv,refs}

\end{document}